%% file: neurips_2026.tex
\documentclass{article}

    \PassOptionsToPackage{numbers, compress}{natbib}
\usepackage[preprint]{neurips_2026}

\usepackage[utf8]{inputenc} 
\usepackage[T1]{fontenc}    
\usepackage{hyperref}       
\usepackage{url}            
\usepackage{booktabs}       
\usepackage{amsfonts}       
\usepackage{nicefrac}       
\usepackage{microtype}      
\usepackage{xcolor}         
\usepackage{graphicx}
\usepackage{amsmath}
\usepackage{wrapfig}
\usepackage{algorithm}
\usepackage[noend]{algpseudocode}
\usepackage{dsfont}

\title{SLayerGen: a Crystal Generative Model for all Space and Layer Groups}

\author{Rees Chang$^1$\thanks{Correspondence to \texttt{reeswc2@illinois.edu}}\ , \textbf{Andrew Novick}$^2$, \textbf{Ryan P. Adams}$^2$, \textbf{Elif Ertekin}$^3$\\
$^1$Department of Materials Science \& Engineering, University of Illinois at Urbana-Champaign\\
$^2$Department of Computer Science, Princeton University\\
$^3$Materials Research Laboratory, University of Illinois at Urbana-Champaign
}

\begin{document}

\maketitle

\begin{abstract}
  Crystal generative models have shown rapid progress for accelerating the discovery of bulk, periodic materials. However, many material systems such as 2D superconductors, thin film semiconductors, and catalytic surfaces are diperiodic, i.e., aperiodic along one of the lattice directions. These systems are invariant under the \emph{layer groups}, which are known to influence materials properties yet not considered by existing models. In this paper, we propose SLayerGen, a generative model that produces crystals constrained to be invariant to any space or layer group. SLayerGen consists of coarse-to-fine discrete autoregressive lattice generation; transformer-based autoregressive sampling of Wyckoff positions, elements, and numbers of symmetrically unique atoms; and space or layer group equivariant diffusion of atomic coordinates. For the diffusion component, we corrected an inconsistency in the loss from prior work arising from hexagonal groups being non-orthogonal in fractional coordinates. To facilitate progress in generative modeling of diperiodic materials, we assembled and filtered datasets of monolayers and bilayers, propose relevant evaluation metrics, and developed novel representations for layer group symmetries. For \emph{de novo} generation of diperiodic materials, SLayerGen achieves consistent performance gains over bulk crystal generative models and is competitive when training jointly on bulk and diperiodic materials.
\end{abstract}

\input{section_introduction}
\input{section_preliminaries}
\input{section_related_works}
\input{section_method}
\input{section_results}
\input{section_conclusion}
\input{section_acknowledgements}

\bibliography{bibliography}
\bibliographystyle{plainnat}
\newpage

\input{section_appendix}


\end{document}

%% file: section_introduction.tex
\section{Introduction}\label{introduction}

Targeted discovery of crystalline materials with functional properties has the potential to drive revolutions in catalysis, computing, energy, and more. Recently, large materials datasets of density functional theory (DFT) calculations \citep{materialsproject, alexandria, omol, omat24, odac25, aqcat25} have led to a large collection of generative models for bulk periodic inorganic crystals, showing great promise to drastically accelerate materials design \citep{mattergen, mofdiff, atommof, mofasa}.

However, generative models have not been developed for materials systems containing both periodic and aperiodic dimensions in 3D space. We focus on diperiodic systems, which span thin film interfaces \citep{intermatch, interface_screening_with_potentials}, surfaces \citep{mp_surface_energies}, monolayers \citep{mofs_2d, cdvae_2d, c2db, c2db_progress, 2dmatpedia, ht_2d_topological}, and multilayer materials \citep{twisted_bilayer_graphene, heterostructures_review, bidb}. Unlike bulk periodic crystals which are invariant to one of 230 discrete \emph{space groups}, diperiodic crystals are invariant to one of 80 discrete \emph{layer groups} (see Sec. \ref{xtal_groups} for definitions).

Diperiodic systems offer large surface to mass ratios, frequently yield exotic physics, and are highly tunable via interlayer stacking, twisting, and sliding. Notably, they have been shown to exhibit symmetry-dependent behaviors such as valley-contrasting physics \citep{valleytronics_2d}, nonlinear optics \citep{nonlinear_optics_2d}, superconductivity \citep{superconductors_2d}, and topological properties \citep{symmetry_breaking_in_2d_materials}, offering potential to unlock new paradigms for flexible electronics, quantum computing, batteries, and more \citep{science_review_2d_materials_and_vdw_hetero}. The layer group symmetries inherent in these systems have been used to theoretically identify bilayer ferroelectrics \citep{layer_groups_2_bilayer_ferroelectrics}, detect 2D topological materials \citep{ht_2d_topological}, and predict nodal semimetallic band structures \citep{semimetals_in_layer_groups}. We highlight that attributing space groups to diperiodic systems by superficially tiling them along the aperiodic direction is inadequate as 18 of the 80 layer groups do not injectively map to a space group  \citep{layer_group_finding_algo}.

\begin{figure*}[!t]
\centering
\includegraphics[width=0.9\linewidth]{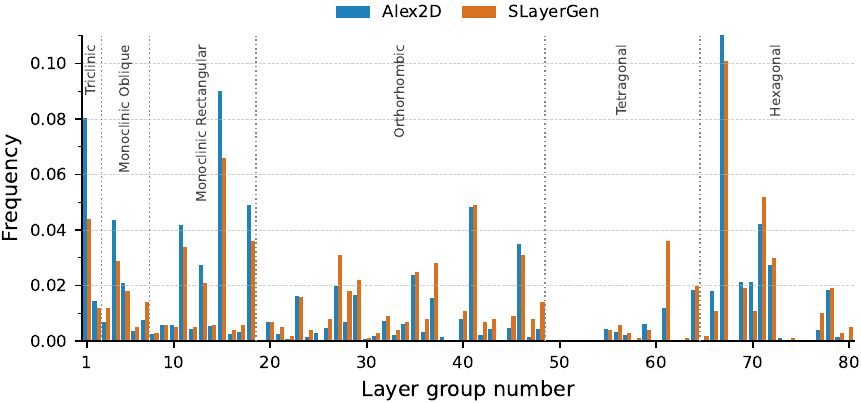}
\vspace{-0.3cm}
\caption{
Histograms of occupied layer groups by samples from SLayerGen and by training data from Alexandria \citep{alexandria, alexandria_2d_strategy}. Layer groups were determined by \texttt{spglib} \citep{layer_group_finding_algo, spglib} using a tolerance of 0.01\AA. Dashed vertical lines separate layer group numbers by crystal system.}
\label{data_figure}
\vspace{-0.5cm}
\end{figure*}

Despite the promise of diperiodic materials, challenges exist for their data-driven design. Firstly, data is scarce. Atomically thin materials were only discovered in 2004 upon the synthesis of graphene \citep{graphene}, and the number of experimentally known mono- or few-layer materials is still $\mathcal{O}(10^2)$ \citep{x2db}. Computational predictions slightly raise dataset sizes to $\mathcal{O}(10^3-10^4)$ after filtering by loose stability metrics \citep{2dmatpedia, c2db_progress, alexandria}. However, dataset construction methods have been limited to elemental substitutions on the few experimentally known structures \citep{mc2d, 2dmatpedia, c2db}, brute-force enumeration through symmetry-derived prototypes \citep{alexandria, alexandria_2d_strategy}, and exfoliability prediction of known bulk periodic crystals \citep{2dmatpedia, ase_detect_2d, cheon_detect_2d, gorai_detect_2d}. For the latter, we note that bulk crystal datasets are not comprehensive \citep{number_of_quaternaries} and atomically thin materials without bulk analogs have been experimentally observed \citep{silicene, 2dmatpedia}. Secondly, layer group symmetry strongly influences materials properties and we empirically observed that databases of diperiodic materials contain a large diversity of layer group symmetries (see Figs. \ref{data_figure} and \ref{all_dataset_layer_hists}). Generative models should thus be capable of reliably reproducing these symmetries. Thirdly, while modeling of periodicity, aperiodicity, continuous translation invariance, and discrete group invariances have been demonstrated across protein, molecular, and bulk crystal generative models \citep{geodiff, torsional_diffusion, protein_torsional_diffusion, cdvae, sgequidiff, space_group_flow_matching}, diperiodic materials uniquely require handling all these attributes in a single model. These challenges demand symmetry-aware AI that can expand and intelligently navigate the space of layer group symmetric materials with high data efficiency.

In this paper, we propose a space and layer group-aware generative model (SLayerGen), which enforces constraints and yields invariant likelihoods with respect to any space group or layer group. SLayerGen combines SE(3)-invariant, coarse-to-fine sampling of discretized lattices; permutation-invariant, transformer-based autogressive generation of occupied \emph{Wyckoff positions} (see Sec. \ref{preliminaries}) and atomic elements; and space or layer group equivariant diffusion of atomic coordinates.

The main contributions of our work are summarized as follows:
\begin{itemize}
    \item We developed embeddings for layer groups and their Wyckoff positions, corrected two layer group asymmetric unit definitions from the International Tables of Crystallography, and enumerated symmetrically unique regions of every Wyckoff position to build uniform priors over symmetrically unique atom positions.
    \item By leveraging components from both bulk crystalline and molecular models, we built the first generative model which, in addition to modeling space group symmetries, explicitly models layer groups. We also corrected an inconsistency between training and model constraints from a prior work when training diffusion models on hexagonal groups.
    \item We curated datasets of monolayers and bilayers to benchmark \emph{de novo} generation of diperiodic materials while proposing several novel evaluation metrics.
    \item We demonstrate state-of-the-art results and competitive performance when jointly training on triperiodic and diperiodic materials.
\end{itemize}

%% file: section_preliminaries.tex
\section{Preliminaries}\label{preliminaries}

\paragraph{Space groups and layer groups}\label{xtal_groups}
All crystals with $d$ periodic dimensions are invariant with respect to at least one crystallographic group $G^d\in\mathbb{G}^d$. A crystallographic group $G^d$ is comprised of an infinite subgroup of discrete lattice translations $T^d_L=\{n_1 L_1, ..., n_dL_d | n_i \in \mathbb{Z}, L_i \in \mathbb{R}^3\}$ and a set of group elements with action $g^d(\cdot) = R_{g^d}(\cdot) + v_{g^d}$ where $R_{g^d}\in\mathrm{O}(3)$ and $v_{g^d}\in\mathbb{R}^3$ is a non-lattice translation. For $d=3$, there are 230 crystallographic \emph{space groups}. For $d=2$, there are 80 crystallographic \emph{layer groups}, generalizing the 17 commonly known wallpaper groups. In the conventional unit cell bases of the layer groups where the center of mass along the $L_3$ axis is set to zero, $v_{g^2}^z=0$ and $R_{g^2}$ is block diagonal with $R_{g^2}=R^{xy}_{g^2} \bigoplus R^z_{g^2}$ where $R^{xy}_{g^2} \in O(2)$ and $R^z_{g^2} \in O(1)$ \citep{layer_group_finding_algo}. For simplicity, we will drop $d$ from the notation going forward and assume equations apply to both $d=2$ and $3$ unless stated otherwise.

\paragraph{Unit cells}
Redundancy from lattice translations $T_L$ can be removed from crystal representation via three basis vectors, $L=[L_1, L_2, L_3]\in\mathbb{R}^{3\times 3}$, which define a parallelepiped $\Gamma \in \mathbb{R}^3$ called the unit cell. A crystal $M$ with $N$ atoms per unit cell can be described by the tuple $M=(A,X,L)$, where $A=(a'_1,...,a'_N)\in \mathbb{A}^N$ are the discrete atomic elements, $X=(Lx'_1,...,Lx'_N)\in\mathbb{R}^{3\times N}$ are the Cartesian atom coordinates, and $\{x'_1,...,x'_N |x'_i\in\Gamma\}$ are the \emph{fractional} atom coordinates. For bulk crystals with space group symmetry, the unit cell is finite and the infinitely periodic crystal can be reconstructed from it as $\{(a'_i,x'_i + n_1e_1+n_2e_2+n_3e_3)_{i=1}^N | n_i \in \mathbb{Z}\}$ where $e_i \in \mathbb{R}^3$ is from the standard basis. In contrast, crystals with layer group symmetry are reconstructed as $\{(a'_i,x'_i + n_1e_1+n_2e_2)_{i=1}^N | n_i \in \mathbb{Z}\}$ and the unit cell is unbounded along $L_3$. Here, $L_3$ merely serves to provide a basis vector for defining atomic coordinates and group representations. In our model, we described atom positions using fractional coordinates for periodic dimensions and Angstrom distances along $L_3$ for aperiodic dimensions. We will refer to the latter as ``aperiodic fractional coordinates''. We canonicalized the non-unique choice of unit cell with the \emph{conventional unit cell} \citep{itc_a} which contains all the symmetries of the crystallographic group.

\paragraph{Wyckoff positions}
Given a crystallographic group and a point $x \in \mathbb{R}^3$, the \emph{stabilizer} group $G_x \equiv \{g|gx=x\} \subset  G$ is the finite subgroup of $G$ that leaves $x$ invariant. A Wyckoff position is the set of points with conjugate stabilizer groups, i.e., $\{x'| \exists\ g\in G :G_{x'}=gG_xg^{-1}\}$. There are 1731 Wyckoff positions across the 230 space groups and 477 Wyckoff positions across the 80 layer groups. When a Wyckoff position's stabilizer group only contains the identity, it is called the \emph{general Wyckoff position} and consists of the interior of the asymmetric unit. When a Wyckoff position's stabilizer group contains nontrivial group elements, it is called a \emph{special Wyckoff position} and either consists of a point (e.g., at an inversion center), set of line segments (e.g., rotation axes), or facets (mirror planes) on the ASU boundaries. The \emph{orbit} size of each Wyckoff position (modulo lattice translations) is called the Wyckoff multiplicity and defined as $|\{gx|g\in G, gx \in \Gamma\}|$.

\paragraph{Asymmetric units}
To enable memory-efficient lazy data loading and construct uniform priors over atom positions that are inequivalent under actions of crystallographic groups, we removed redundancy induced by the crystallographic group $G \supseteq T_L$ by storing crystals in terms of their \emph{asymmetric unit} (ASU) $\Pi\in\mathbb{R}^3$. In this representation, a crystal is described as fractional coordinates $X=\{(x_1, ...,x_n)|x_i\in\Pi\}\in \mathbb{R}^{3\times n}$, atomic elements $A=(a_1,...,a_n)\in\mathbb{A}^n$, and Wyckoff positions $W=\{(w_1,...,w_n)|w_i=G_{x_i}\}\in\mathbb{W}^n$ where $n \leq N$. The infinitely periodic crystal can be reconstructed from the asymmetric unit as $\{(a_i,g_{ij}x_i)\ |\ x_i\in \Pi, \ g_{ij} \in G/w_i,\ i\in (1, ..., n)\}$. We canonicalized the non-unique choice of ASU using those in the International Tables for Crystallography \citep{itc_a}. To ensure that the ASU is \emph{exact}, i.e., tiles $\mathbb{R}^3$ without overlaps at boundaries, we additionally defined which ASU faces, edges, and vertices are (partially) open or closed (see Sec. \ref{crystal_representation}).

\paragraph{Crystallographic group equivariant neural networks}
Given a group of isometries $G'$, \citet{rezende2019equivarianthamiltonianflows} and \citet{kohler20equivariantflows} showed that transforming a $G'$-invariant probability density according to a $G'$-equivariant flow will preserve invariance. \citet{Yarotsky2022} and \citet{puny_frame_averaging} then showed how to symmetrize neural networks to be invariant or equivariant via group averaging without reducing expressivity of the original model. Symmetrization was later used to build space group equivariant functions $f:\mathbb{R}^3\rightarrow \mathbb{R}^3$ for bulk crystals and metamaterials \citep{sgequidiff, space_group_flow_matching, space_group_equivariant_neural_networks} given a $T_L$-invariant neural network $\hat{f}: \mathbb{R}^3 \rightarrow \mathbb{R}^3$ using the following operator:
\begin{align} \label{sg_equivariance}
    f(x) = \frac{1}{|G/T_L|} \sum_{g\in G/T_L} R^{-1}_g \hat{f}(gx).
\end{align}
Eq. \ref{sg_equivariance} has been proven to satisfy equivariance, i.e., $f(gx)=R_gf(x)$. We use the same operation to symmetrize a denoising score network to be equivariant with respect to any space or layer group. 

\citet{sgequidiff} and \citet{space_group_flow_matching} also showed that transforming atom positions according to a space group equivariant flow preserves the stabilizer group (i.e., Wyckoff position) of the atoms. Specifically, the following holds for arbitrary scalar $\epsilon \in \mathbb{R}$ and stabilizer group element $g_x \in G_x$:
\begin{align}\label{stabilizer_preservation}
    g_x(x+\epsilon f(x))=x+\epsilon f(x) \implies G_x \subseteq G_{x+\epsilon f(x)}.
\end{align}

%% file: section_related_works.tex
\section{Related Work}

\paragraph{Protein folding and molecular generation} Here, we give some relevant examples of generative models for molecules and proteins. Early generative models for molecules focused on discrete string representations \citep{rafa_molecule_vae}, later generating all-atom 3D geometries autoregressively \citep{gschnet} and with E(3)-equivariant diffusion models \citep{confgf, geodiff, e3_molecular_diffusion}. Other approaches diffused relative atom orientations, e.g., via periodic torsion angles \citep{torsional_diffusion, protein_torsional_diffusion}. Similarly to molecules and proteins, diperiodic materials require generation of periodic and continuously translation invariant aperiodic coordinates. However, diperiodic materials also have invariances to non-translational discrete isometries.

\paragraph{Generative models for bulk materials}
In the last few years, a large body of generative models have been developed for bulk crystalline materials, including variational autoencoders (VAE) \citep{ftcp, gaussian_voxels_crystal_vae, hoffman, cdvae}, generative adversarial networks \citep{gan_on_cell_parameters_with_data_augmentation}, diffusion models \citep{cdvae, diffcsp, mattergen, equicsp, unimat, joshi2025allatom}, flow matching models \citep{flowmm}, stochastic interpolants \citep{omg_crystal_stochastic_interpolants}, Bayesian Flow Networks \citep{crysbfn}, and fine-tuned large language models \citep{flowllm, genms, chemeleon, finetuned_crystal_llm, invariant_llm_tokenization}.

Several models have been developed which consider space group constraints during generation. Some of these models focused on generation of discrete space groups and Wyckoff positions \citep{wycryst, crystalgfn, diffcsp_pp, crystalformer, kazeev2024wyckofftransformer, wyckoffdiff}, while others focused on constrained generation of atom coordinates using P1-invariant models and projections onto Wyckoff positions \citep{diffcsp_pp, symmcd, symmbfn}. Recently, SGEquiDiff \citep{sgequidiff} and SGFM \citep{space_group_flow_matching} constructed space group invariant model likelihoods and achieved state-of-the-art performance by symmetrizing diffusion and flow matching models with space group equivariance, automatically preserving space group symmetries at every diffusion/integration step. SLayerGen builds on the work of SGEquiDiff to enable layer group constrained generative modeling for the first time while preserving the ability to model space groups.

\paragraph{Data-driven prediction of diperiodic materials} Besides the aforementioned methods for diperiodic materials dataset construction (Sec. \ref{introduction}), several data-driven approaches have been applied to diperiodic materials design. Some works trained crystal diffusion models that spuriously assume diperiodic materials have triperiodicity \citep{cdvae_2d, cdvae_heterostructures, space_group_2d_generative}. Another work \citep{composition_2d_transformer} predicted elemental compositions with a transformer, selected structural prototypes from known diperiodic materials, and relaxed atom positions with a machine learning interatomic potential (MLIP) \citep{composition_2d_transformer}. Unlike these approaches, SLayerGen models the layer group symmetries of diperiodic materials and the space group symmetries of bulk materials while generating entire atomic structures from scratch.

%% file: section_method.tex
\section{SLayerGen}

\begin{figure*}[!t]
\centering
\includegraphics[width=\textwidth]{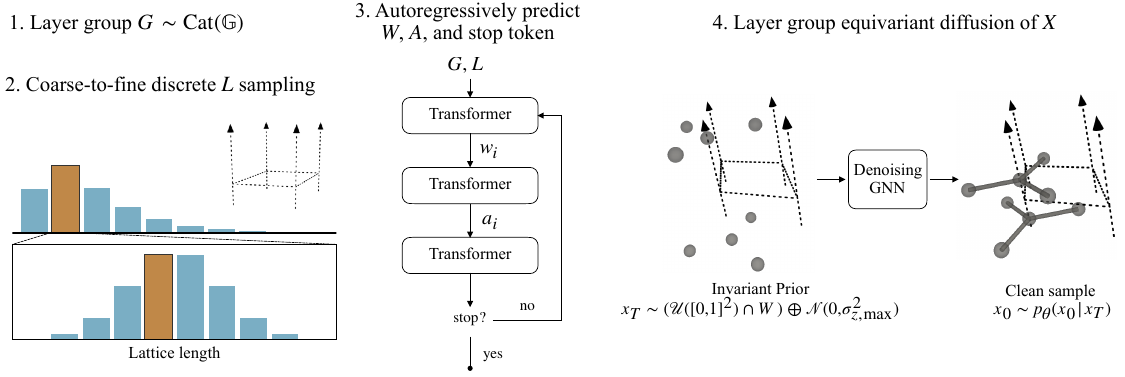}
\vspace{-0.7cm}
\caption{Illustration of our diperiodic crystal generation process.}
\label{generation_process}
\vspace{-0.5cm}
\end{figure*}

\subsection{Crystal representation and generation process}\label{crystal_representation}
We represented each crystal as a tuple $M=(G, L, A, W, X)\in(\mathbb{G}, \mathbb{R}^{3\times 3}, \mathbb{A}^n, \mathbb{W}^n, \mathbb{R}^{n\times 3})$, where $n$ is the number of atoms in the exact asymmetric unit, $G\in \mathbb{G}$ is the space or layer group, $L\in \mathbb{R}^{3\times 3}$ is the conventional lattice basis, $A\in \mathbb{A}^n$ are elements, $W\in \mathbb{W}^n$ are Wyckoff positions, and $X\in\mathbb{R}^{n\times 3}$ are fractional atom coordinates. To handle directional dependencies from symmetry constraints, we factorized generation as follows:
\begin{align} \label{xtal_factorization}
    p(M)= p(G)\times p(L|G)\times p(W,A|L,G) \times p(X|W,A,L,G).
\end{align}
To reduce memory requirements and enable atom diffusion from uniform priors over symmetrically inequivalent atom positions, we removed redundancy from each Wyckoff position by finding its intersection with the corresponding exact ASU. We refer to these intersections as \emph{Wyckoff shapes} as they consisted of points, lines, line segments, planes, polygons, and interiors of the ASUs. For the space groups, we used the intersections enumerated by SGEquiDiff \citep{sgequidiff} which leveraged exact ASU boundary conditions found by \citet{exact_asus}. However, exact ASUs for the layer groups have not been enumerated to the best of our knowledge, so we removed redundancy from every layer group Wyckoff position ourselves using custom visualization and validation scripts (see Sec. \ref{layer_asus} for details). During this process, we found and corrected erroneous entries in the inexact asymmetric units listed for layer groups 7 and 46 in the International Tables of Crystallography \citep{itc_e}.

We set $p(G)$ to the empirical training distribution. Now we discuss the remaining distributions.

\subsection{Coarse-to-fine discrete sampling of $L$}
Most crystallographic groups impose constraints on the crystal lattice lengths $(a=||L_1||_2,b=||L_2||_2,c=||L_3||_2)$ and angles $(\alpha=\angle L_2L_3,\beta=\angle L_1L_3,\gamma=\angle L_1L_2)$ (see Sec. \ref{space_group_lattices}). To enforce these constraints during lattice generation in SLayerGen, we extended the lattice generation model of SGEquiDiff \citep{sgequidiff} to support layer group constraints and to mask out generation of the superficial lattice length $c$. While DiffCSP++ \citep{diffcsp_pp} also proposed a masked diffusion approach for constraining lattice parameters, it does not enforce positive determinant in the lattice matrix and thus is not guaranteed to preserve the 22 chiral space groups with enantiomorphic pairs \citep{itc_a}. Briefly, our model achieves efficient high-resolution and constrained lattice generation via coarse-to-fine autoregressive sampling of discretized lattice parameters where crystallographic group constraints are enforced at training and inference time via logit masking. The model likelihoods are modeled as follows:
\begin{equation*}\label{lattice_eq1}
p(L|G)=\prod\limits_{i=1}^6 p(l_i|l_{<i},G) \qquad p(l_i|l_{<i},G)=\prod^K_{k=1}p(b_k|b_{k-1},l_{<i}, G)
\end{equation*}
where $l_i\in\mathbb{R}$ is a lattice parameter, $K$ is the number of resolutions, and $b_k$ is a class from the categorical distribution at resolution $k$ with $n_l$ categories. We trained an MLP with the cross-entropy loss with $K=2$ and $n_l=100$. Lattice parameters were modeled as the sequence, ($a$, $b$, $c$, $\alpha$, $\beta$, $\gamma$), and we maximized likelihoods of ground truth parameters conditioned on preceding lattice lengths and angles noised with uniform random noise with 0.3\AA\  and 5$^\circ$ ranges, respectively.

\subsection{Transformer-based autoregressive sampling of $W$, $A$, and $n$}
We used a transformer based on SGEquiDiff \citep{sgequidiff} to autoregressively generate Wyckoff positions and elements. A learnable stop token was used to implicitly learn the number of atoms per material. Sequences of Wyckoff positions and elements were ordered lexicographically and logits of Wyckoff positions consisting of 0D points were masked if atoms were already generated there. The following factorized model likelihood was trained with a cross-entropy loss:
\begin{align}
	p(W,A|L,G) &= \prod_{i=1}^n\Big[ p(w_i|w_{<i},a_{<i},L,G) p(a_i|w_{\leq i},a_{<i},L,G) p(\mathrm{stop}|w_{\leq i},a_{\leq i},L,G) \Big].
\end{align}
To extend SGEquiDiff \citep{sgequidiff} for layer groups and encourage learning correlations across different groups, we created embeddings of layer groups and their Wyckoff positions consistently with SGEquiDiff's embeddings for space groups and their Wyckoff positions. These embeddings were used to encode intermediate states and keys in the transformer. See Sec. \ref{layer_group_wyckoff_embeddings} for details.

\subsection{Space and layer group equivariant diffusion of $X$}
    We generated atomic fractional coordinates with noise conditional score matching diffusion \citep{ncsn}. Similarly to SGEquiDiff \citep{sgequidiff}, we learned scores with a $G$-invariant "group wrapped normal'' (GWN):
    \begin{align}\label{gwn}
    p(x_t|x_0) &\propto \sum_{g\in G} \exp\Big(-\frac{1}{2}(x_t - g x_0)^T \Sigma_t^{-1} (x_t - gx_0)\Big)
    \end{align}
    where $x_t$ is a noisy sample, $x_0$ is a clean sample, and $\Sigma_t$ is the noise covariance matrix at time $t$. We made several changes compared to SGEquiDiff. First, since layer groups are invariant with respect to continuous global translations along the aperiodic direction, we followed work on molecular and particle systems \citep{geodiff, kohler20equivariantflows} to build \emph{center of mass (CoM)-free distributions}; we always translated the mean $z$-coordinate of our diperiodic materials to zero, including rigidly aligning the $z$-component of the CoM between clean and noised diperiodic structures before score computation. Second, when $G$ is a layer group and the noise scale $\Sigma_t$ increases, the distribution does not approach uniform in the aperiodic direction as it does in the periodic directions. To handle this, we chose diagonal but anisotropic noise covariance $\Sigma_t = \sigma^2_{t,xy}I_2 \bigoplus \sigma^2_{t,z} I_1$ where $\sigma_{t,xy},\sigma_{t,z} \in \mathbb{R}$, $I_d \in \mathbb{R}^{d\times d}$ is the identity, and $\sigma_{t,z}$ describes the noise on aperiodic coordinates in Angstroms. Using variance exploding noise schedules, we note that the resulting noise prior, uniform in the periodic plane ($\mathcal{U}([0, 1]^2)$) and Gaussian along the $L_3$-axis ($\mathcal{N}(0, \sigma^2_{T,z})$), is trivially invariant to all layer groups. Third, \citet{sgequidiff} proved that the score of a GWN with isotropic noise is equivariant, i.e., $\nabla_{x_t}\log p(gx_t|x_0) = R_g\nabla_{x_t}\log p(x_t|x_0)$, and thus preserves invariant distributions \citep{kohler20equivariantflows} and stabilizer groups via Eq. \ref{stabilizer_preservation}. We prove equivariance also holds for layer groups with our choice of $\Sigma_t$ in Sec. \ref{layer_score_equivariance}. 
    
    The equivariance proof by \citet{sgequidiff} requires that the matrix representation $R_g$, where $g=\{R_g, v_g\}\in G$, is orthogonal; however, while $R_g$ is always orthogonal in real space, it is not orthogonal \emph{in fractional coordinates} for hexagonal groups, as noted by \citet{space_group_flow_matching}. We verified empirically that ignoring the non-orthogonality results in non-equivariant GWN scores in fractional coordinates, yielding a mismatch between target scores and predictions from models symmetrized by Eq. \ref{sg_equivariance}. To resolve this issue, we computed scores of hexagonal groups in a dimensionless Cartesian basis and then converted the scores back to fractional coordinates. In Sec. \ref{hexagonal_scores}, we describe the induced score function $\phi(x_t|x_0)$ in fractional coordinates and prove that it satisfies equivariance, i.e., $\phi(gx_t|x_0)=R_g\phi(x_t|x_0)$.

    We learned the scores of Eq. \ref{gwn} by reconstructing the conventional unit cell from its asymmetric unit and symmetrizing a denoising graph neural network using Eq. \ref{sg_equivariance}. Plane waves and random Fourier features were used to embed periodic and aperiodic coordinates, respectively. See Sec. \ref{gnn_architecture} and \ref{diffusion_details} for further details on the diffusion model architecture, training, and inference.

%% file: section_results.tex
\section{Experiments}
    \paragraph{Data processing} Since only a few hundred diperiodic materials have been experimentally cataloged \citep{x2db}, we evaluated SLayerGen on four DFT-based datasets of diperiodic materials: 2DMatPedia \citep{2dmatpedia}, C2DB \citep{c2db}, BiDB \citep{bidb}, and Alex2D \citep{alexandria, alexandria_2d_strategy}. 2DMatPedia was constructed by screening all layered bulk materials (including hypothetical ones) from the Materials Project \citep{materialsproject}, theoretically exfoliating them into monolayers, and substituting their elements. C2DB conducted elemental substitutions only in experimentally known monolayers. BiDB extracted stable monolayers from C2DB with up to 10 atoms per unit cell and stacked them in various configurations to create homobilayers. Alex2D enumerated hypothetical prototypes of layer groups and Wyckoff positions and then occupied them with elements that satisfy charge neutrality. We note that merging datasets is nontrivial as each one uses distinct DFT settings (see Sec. \ref{dataset_settings}).

    \input{evals_2dmatpedia}
    \input{evals_c2db}

    We filtered materials from each dataset that are unlikely to be synthesizable. Based on an analysis of experimentally known monolayers \citep{c2db}, structures from C2DB and Alex2D were removed if their reported DFT energies above the hull were greater than 200 meV/atom. Since 2DMatPedia used a different DFT functional (vdW-optB88), we filtered that dataset with a cutoff of 150 meV/atom, also determined from known monolayers \citep{2dmatpedia}. We filtered BiDB bilayers based on their reported thermodynamic binding stabilities with respect to isolated layers ($E_b = E_\mathrm{bilayer} - 2E_\mathrm{monolayer} < 0$) and dynamical stabilities with respect to interlayer sliding.
    
    All datasets were further filtered by removing structures predicted to be dynamically unstable with respect to lattice commensurate vibrations, i.e., $\Gamma$-point phonons. Dynamical stability was determined by the finite displacement method \citep{finite_displacement_phonons} under the harmonic approximation with a foundation machine learning interatomic potential (MLIP) accelerated by \texttt{torchsim} \citep{torchsim}. Specifically, we used \texttt{orb\_v3\_conservative\_inf\_mpa} \citep{orbv3} for 2DMatPedia, C2DB, and Alex2D, and \texttt{orb\_v3\_conservative\_inf\_omat} with BJ-damped D3 corrections \citep{bj_damped_d3} for BiDB to account for interlayer van der Waals forces. We highlight that Orb v3 was found to yield state-of-the-art phonon frequency predictions (MAEs <5 meV compared to PBE DFT calculations) on a benchmark of 5000 stable crystals covering 86 elements \citep{orbv3_good_at_phonons}, and its predecessor Orb v2 was found to excel at structure optimization in heterobilayers \citep{mlip_heterobilayers}. We verified these claims by calculating F1 scores for Orb v3 when predicting thermodynamic stability on these datasets, finding significantly higher performance than a dummy classifier that predicts everything is stable (Table \ref{orb_f1_scores}). Following \citet{mlips_are_ready_for_phonons}, we considered a structure dynamically stable with respect to $\Gamma$-point phonons if all frequencies were real except for the three acoustic modes, which were allowed to have small imaginary frequencies of -50 K to account for numerical errors.
    
    We note that our filtering tolerances are purposely looser than typical for bulk crystals \citep{wenhao_metastability, cdvae}; the convex hulls here merely serve as a rough guide for the stability of diperiodic materials as surface and interfacial interactions were neglected. Correspondingly, many free-standing diperiodic structures with large energies above the hull and/or unstable phonons have been experimentally realized using substrate materials \citep{2dmatpedia, silicene, germanene, moO3}, metal ion intercalation \citep{metal_atom_intercalation}, or self-assembly in solution \citep{lb_graphene}.
    
    After data filtering, we constructed 70/15/15 data splits from the remaining 2,696 of 6,351 monolayers from 2DMatPedia, 8,502 of 16,905 monolayers from C2DB, 2,641 of 11,184 bilayers from BiDB, and 30,082 of 137,833 materials from Alex2D. We standardized all data and generated structures, superficially represented as having bounded unit cells, by padding the ``lattice length'' in the aperiodic direction to 25\AA\ plus the max empirical layer thickness and then centering the mean of each layer to $z=0.5$ (see Sec. \ref{layer_centering}). Layer groups and Wyckoff positions were determined with \texttt{spglib.get\_layergroup} using \texttt{symprec=0.01} for data pre-processing and \texttt{symprec=0.1} for model evaluation \citep{spglib, layer_group_finding_algo}.

    To assess whether SLayerGen's ability to handle layer groups harms its performance on space group symmetric materials, we also provide results on the MP20 dataset \citep{cdvae}, containing 45,231 experimentally observed triperiodic crystals with up to 20 atoms per unit cell.

    \paragraph{Baselines} DiffCSP (12.3M parameters) \citep{diffcsp} enforces P1 space group symmetry and jointly diffuses elements, lattice matrices, and fractional atom coordinates. The model uses a P1-invariant GNN with fully-connected graphs under periodicity. MatterGen (44.6M parameters) \citep{mattergen} similarly enforces P1 space groups and does joint diffusion, but leverages Cartesian coordinates and an SE(3)-equivariant GNN with radius-based graphs. DiffCSP++ \citep{diffcsp_pp} enforces constraints from larger space groups with per-step projections during diffusion given templates from training data. SGEquiDiff (5.5M parameters) \citep{sgequidiff} also enforces space group constraints but builds crystals from scratch by combining autoregressive sampling of lattices, Wyckoff positions, and elements with space group equivariant diffusion of atom coordinates. We note that DiffCSP, DiffCSP++, and MatterGen do not learn the distribution over number of atoms per unit cell and DiffCSP++ does not learn the distribution over occupied Wyckoff positions, instead relying on examples from training data. 
    
    To isolate the effect of enforcing layer group symmetries from model architecture, training recipe, and other hyperparameters, we also compared to \emph{SLayerGen3D}, which is identical to SLayerGen (6.0M parameters) but trains on diperiodic materials represented as having space group symmetries.

    \input{evals_bidb}
    \input{evals_alex2d}
    \input{evals_mp20}

    \paragraph{Metrics} Each generative model was evaluated on a budget of 10,000 generated materials. We computed average sampling times per batch of 500 materials on a single A40 NVIDIA GPU, structural validity defined by all atomic nearest neighbor distances being between 0.5\AA\ and 7.5\AA, and compositional validity based on \texttt{SMACT} \citep{number_of_quaternaries}. We note that roughly 10\% of known crystals fail the compositional validity checker \citep{cdvae}. Of 1000 structurally and compositionally valid random samples, we measured goodness-of-fit via distributional distances to the test set: Jensen-Shannon divergences (JSD) for the layer groups and Wasserstein distances for areal densities projected along the lattice plane $\rho$ ($\mu$g/cm$^2$), monolayer thicknesses (\AA), bilayer interlayer distances (\AA), and numbers of unique elements per structure $N_\mathrm{el}$. We then computed uniqueness and novelty (U.N.) rates with two methods. First, we measured U.N. \emph{Structure} rates with respect to the training set using \texttt{pymatgen StructureMatcher} \citep{pymatgen} (\texttt{ltol=0.2}, \texttt{stol=0.3}, and \texttt{angle\_tol=5}), which applies affine lattice transformations to match atomic structures within user tolerances. Second, following SymmCD \citep{symmcd}, we computed chemistry-agnostic U.N. \emph{template} rates, where a template was defined as the multiset of a structure's layer group and occupied Wyckoff positions.

    To assess the ability of generative models to find new realistic structures, we conducted relaxations on 1,000 valid structures per model using \texttt{Orb-v3} \citep{orbv3}; we then measured stability and recomputed uniqueness and novelty based on the relaxed structures. For the BiDB dataset, we used \texttt{orb\_v3\_conservative\_inf\_omat} with BJ-dampled D3 corrections. Stability was defined as the relaxed bilayer having lower total energy than its individual relaxed monolayers (B) and having dynamically stable $\Gamma$-point phonons. For all other datasets, we used \texttt{orb\_v3\_conservative\_inf\_mpa} and defined stability as having (1) an energy above the hull $\mathrm{E_{hull}}$ less than 200 meV/atom with respect to the Materials Project \texttt{v2023.02.07} and (2) dynamically stable $\Gamma$-point phonons. We report stability rates with and without the phonon constraint as well as the rate of S.U.N. structures and templates, where we introduce the S.U.N. template rate to assess how often a stable structure with a unique and novel template is generated. This metric aims to measure generalization beyond traditional elemental substitutions in known structures.
    
    MLIP-based relaxations were conducted with a force convergence tolerance of 0.005 eV/\AA\ using the fast inertial relaxation engine \citep{fire_algo}. To prevent spurious interactions between periodic images along the aperiodic direction, we padded the lattice $c$-axis to 25\AA\ of vacuum and fixed it during relaxation. Before computing uniqueness or novelty of structures, they were all padded to the same $c$-axis length as \texttt{StructureMatcher} is sensitive to variable lengths.

    \paragraph{Results} On the 2DMatPedia, C2DB, and Alex2D datasets, SLayerGen achieved the best JSD metric for layer groups, including compared to space group constrained models (Tables \ref{2dmatpedia_results}-\ref{alex2d_results}). While DiffCSP++ beat SLayerGen's layer group JSD metric on the BiDB dataset, we attribute the difference to DiffCSP++ leveraging ground truth combinations of Wyckoff positions from training data. In Fig. \ref{all_model_layer_hists}, we show histograms of the layer groups of structures generated by MatterGen, DiffCSP++, and SLayerGen after training on Alex2D. The distributions reveal that MatterGen heavily oversamples P1 crystals (potentially explaining its inflated (S.)U.N. structure rates since P1 groups impose the fewest constraints on atom positions) and DiffCSP++ cannot sample many monoclinic or orthorhombic layer groups, while SLayerGen closely reproduces the training distribution of layer groups.

    Significantly, SLayerGen also achieved the highest S.U.N. template rate on every dataset with the fastest or second fastest generation speed, highlighting SLayerGen's ability to generalize to unseen structure templates. We show examples of SLayerGen-generated diperiodic materials with S.U.N. templates in Fig. \ref{alex2d_sun_templates}. While MatterGen had a higher S.U.N. template rate than the default SLayerGen configuration on the Alex2D dataset, SLayerGen's modular architecture permits selective inference-time tempering. We leveraged this capability to temper the distributions over layer groups and Wyckoff positions (SLayerGen $T_{G,W}=2.0$), leading to the highest S.U.N. template rate at the cost of slightly lower stability and S.U.N. structure rates (Table \ref{alex2d_results}).

    By comparing SLayerGen to SLayerGen3D, we found that modeling diperiodic materials with layer groups instead of space groups consistently enhanced performance. In particular, SLayerGen always obtained lower distributional distances for layer groups and areal densities, and most importantly, higher S.U.N. template rates. On 3 of the 4 datasets, SLayerGen also achieved better validity rates, stability rates, S.U.N. structure rates, and distributional distances for thickness.

    Since SLayerGen can condition on any space or layer group, we evaluated the model's ability to train on both di- and triperiodic crystals. On most metrics for C2DB and MP20, SLayerGen3D performed comparably or slightly better than SGEquiDiff, the baseline model upon which SLayerGen is most closely related (Tables \ref{c2db_results} and \ref{mp20_results}). When training SLayerGen jointly on Alex2D and MP20, we observed higher U.N. and S.U.N. rates on MP20 (Table \ref{mp20_results}) and nearly identical performance to SLayerGen when only trained on Alex2D (Table \ref{alex2d_results}).

    Beyond model comparisons, we found that dynamical $\Gamma$ phonon stability, ignored in prior generative model evaluations \citep{mattergen, flowmm, sgequidiff}, significantly decreased stability rates.

%% file: evals_2dmatpedia.tex
\begin{table}[t]
\centering
\caption{Results on the 2DMatPedia dataset. Best value per metric is bold, second best is underlined.}
\vspace{-0.2cm}
\resizebox{\textwidth}{!}{
\begin{tabular}{cccccccccccccc}
\midrule
& \multicolumn{1}{c}{Time $\downarrow$} & \multicolumn{2}{c}{Validity (\%) $\uparrow$} & \multicolumn{2}{c}{U.N. rate (\%) $\uparrow$} & \multicolumn{4}{c}{Distribution distance $\downarrow$} & \multicolumn{2}{c}{Stable (\%) $\uparrow$}& \multicolumn{2}{c}{S.U.N. (\%) $\uparrow$}\\
 & (s / batch) & Structure & Composition & Structure & Template & $W_\mathrm{Thickness}$ & $W_\rho$ & $W_{N_\mathrm{el}}$ & $\mathrm{JSD}_G$ & $\mathrm{E_{hull}}$ & $\mathrm{E_{hull}}\Gamma$ & Structure & Template\\ \midrule
DiffCSP & 212 & 94.9 & \textbf{82.3} & \underline{98.2} & 9.4 & 1.054 & 0.052 & 3.373 & 0.762 & 17.5 & 11.7 & 10.6 & 2.4\\
DiffCSP++  & 247 & 96.0 & 73.4 & 94.2 & 9.9 & \textbf{0.471} & 0.061 & 1.447 & 0.350 & 31.1 & 21.1 & 15.8 & 2.9\\
MatterGen  & 1555 & \textbf{99.0} & \underline{78.6} & \textbf{98.8} & 9.8 & 5.118 & \textbf{0.036} & 1.858 & 0.701 & \underline{41.5} & \textbf{33.2} & \textbf{30.7} & \underline{4.2}\\
\midrule
SLayerGen3D & \underline{149} & 93.5 & 71.8 & 92.4 & \underline{23.6} & 1.301 & 0.098 & \textbf{0.147} & \underline{0.348} & 32.8 & 25.6 & 20.3 & 3.6\\\midrule
SLayerGen  & \textbf{143} & \underline{98.3} & 70.9 & 89.1 & \textbf{29.6} & 1.169 & \underline{0.044} & \underline{0.185} & \textbf{0.194} & \textbf{43.8} & \textbf{33.2} & \underline{24.8} & \textbf{5.4}\\
SLayerGen ($T_{G,W}=2.0$)  & \textbf{136} & 96.8 & 68.1 & 94.1 & \textbf{48.0} & \underline{1.040} & \underline{0.044} & \underline{0.255} & \textbf{0.319} & 37.7 & \underline{26.5} & 21.5 & \textbf{7.0}\\
\midrule
\end{tabular}}
\label{2dmatpedia_results}
\vspace{-0.4cm}
\end{table}

%% file: evals_c2db.tex
\begin{table}[t]
\centering
\caption{Results on the C2DB dataset. Best value per metric is bold, second best is underlined.}
\vspace{-0.2cm}
\resizebox{\textwidth}{!}{
\begin{tabular}{cccccccccccccc}
\midrule
& \multicolumn{1}{c}{Time $\downarrow$} & \multicolumn{2}{c}{Validity (\%) $\uparrow$} & \multicolumn{2}{c}{U.N. rate (\%) $\uparrow$} & \multicolumn{4}{c}{Distribution distance $\downarrow$} & \multicolumn{2}{c}{Stable (\%) $\uparrow$}& \multicolumn{2}{c}{S.U.N. (\%) $\uparrow$}\\
 & (s / batch) & Structure & Composition & Structure & Template & $W_\mathrm{Thickness}$ & $W_\rho$ & $W_{N_\mathrm{el}}$ & $\mathrm{JSD}_G$ & $\mathrm{E_{hull}}$ & $\mathrm{E_{hull}}\Gamma$ & Structure & Template\\ \midrule
DiffCSP & 126 & 99.2 & 82.6 & \underline{89.0} & 13.0 & \underline{0.265} & \textbf{0.022} & 1.313 & 0.617 & 52.7 & 49.4 & \underline{35.5} & 4.3\\
DiffCSP++  & 141 & \underline{99.6} & \underline{83.4} & 77.4 & 7.4 & \textbf{0.236} & 0.057 & 0.579 & 0.311 & 58.9 & 49.7 & 32.2 & 4.4\\
MatterGen & 1308 & \textbf{99.7} & 80.2 & \textbf{94.3} & 10.1 & 0.508 & 0.045 & 0.648 & 0.521 & \underline{59.3} & \textbf{54.3} & \textbf{45.2} & \underline{5.0}\\
SGEquiDiff   & \underline{82} & 96.7 & 83.2 & 80.4 & \underline{16.1} & 1.098 & 0.061 & 0.317 & 0.307 & 42.7 & 36.3 & 20.3 & 3.1\\ \midrule
SLayerGen3D & 94 & 99.0 & 83.2 & 71.6 & 15.6 & 0.384 & 0.087 & \underline{0.109} & \underline{0.295} & 58.3 & 50.4 & 24.7 & 3.4\\\midrule
SLayerGen  & \textbf{81} & \textbf{99.7} & \textbf{84.4} & 76.5 & \textbf{19.9} & 0.604 & \underline{0.037} & 0.173 & \textbf{0.183} & \textbf{60.0} & \underline{54.1} & 33.0 & \textbf{5.6}\\
SLayerGen ($T_{G,W}=2.0$)  & 86 & 99.2 & 81.0 & 83.6 & \textbf{38.4} & 0.791 & 0.050 & \textbf{0.069} & \textbf{0.239} & 54.3 & 46.7 & 30.0 & \textbf{6.8}\\
\midrule
\end{tabular}}
\label{c2db_results}
\vspace{-0.7cm}
\end{table}

%% file: evals_bidb.tex
\begin{table}[t]
\centering
\caption{Results on the BiDB dataset. Best result is bold, second best is underlined.}
\vspace{-0.2cm}
\resizebox{\textwidth}{!}{
\begin{tabular}{ccccccccccccccc}
\midrule
& \multicolumn{1}{c}{Time $\downarrow$} & \multicolumn{2}{c}{Validity (\%) $\uparrow$} & \multicolumn{2}{c}{U.N. rate (\%) $\uparrow$} & \multicolumn{5}{c}{Distribution distance $\downarrow$} & \multicolumn{2}{c}{Stable (\%) $\uparrow$} & \multicolumn{2}{c}{S.U.N. (\%) $\uparrow$}\\
 & (s / batch) & Structure & Composition & Structure & Template & $W_\mathrm{Thickness}$ & $W_{d_\mathrm{Interlayer}}$ & $W_\rho$ & $W_{N_\mathrm{el}}$ & $\mathrm{JSD}_G$ & B & B$\Gamma$ & Structure & Template\\ \midrule
DiffCSP & 158 & \textbf{99.6} & 79.9 & \textbf{96.8} & 10.6 & 0.188 & \underline{0.036} & \underline{0.312} & 2.050 & 0.597 & 80.0 & 62.8 & \textbf{63.6} & 4.7\\
DiffCSP++  & 221 & 97.3 & \underline{83.3} & 78.6 & 2.4 & \underline{0.094} & \textbf{0.020} & \textbf{0.242} & 0.738 & \textbf{0.134} & 89.7 & 70.3 & \underline{52.6} & 4.6\\
MatterGen & 1896 & \textbf{99.6} & 82.0 & \underline{79.4} & 8.9 & 0.324 & 0.043 & 0.483 & 0.655 & 0.411 & \underline{91.0} & \underline{80.6} & 51.2 & 5.0\\
\midrule
SLayerGen3D   & \textbf{116} & \underline{98.7} & 82.4 & 57.8 & \underline{12.2} & 0.163 & 0.060 & 0.843 & \underline{0.312} & 0.213 & \textbf{93.5} & \textbf{80.7} & 41.8 & \underline{7.3}\\ \midrule
SLayerGen   & \underline{131} & 97.8 & \textbf{86.1} & 50.3 & 10.9 & \textbf{0.061} & 0.044 & 0.638 & \textbf{0.188} & \underline{0.178} & \underline{93.2} & 80.2 & 36.2 & \textbf{7.6}\\
SLayerGen ($T_{G,W}=2.0$)  & 164 & 95.9 & 80.4 & 60.6 & \textbf{22.5} & 0.160 & 0.085 & 0.998 & \textbf{0.209} & 0.268 & \textbf{93.5} & 73.4 & 37.2 & \textbf{11.3}\\
\midrule
\end{tabular}}
\label{bidb_results}
\vspace{-0.35cm}
\end{table}

%% file: evals_alex2d.tex
\begin{table}[t]
\centering
\caption{Results on the Alex2D dataset. Best result is bold, second best is underlined.}
\vspace{-0.2cm}
\resizebox{\textwidth}{!}{
\begin{tabular}{cccccccccccccc}
\midrule
& \multicolumn{1}{c}{Time $\downarrow$} & \multicolumn{2}{c}{Validity (\%) $\uparrow$} & \multicolumn{2}{c}{U.N. rate (\%) $\uparrow$} & \multicolumn{4}{c}{Distribution distance $\downarrow$} & \multicolumn{2}{c}{Stable (\%) $\uparrow$}& \multicolumn{2}{c}{S.U.N. (\%) $\uparrow$}\\
 & (s / batch) & Structure & Composition & Structure & Template & $W_\mathrm{Thickness}$ & $W_\rho$ & $W_{N_\mathrm{el}}$ & $\mathrm{JSD}_G$ & $\mathrm{E_{hull}}$ & $\mathrm{E_{hull}}\Gamma$ & Structure & Template\\ \midrule
DiffCSP  & 125 & 99.1 & 72.7 & \textbf{87.8} & 7.3 & \textbf{0.130} & \underline{0.015} & 1.072 & 0.614 & 59.7 & 52.9 & \underline{45.1} & 2.5\\
DiffCSP++  & 137 & \textbf{99.9} & 75.8 & 70.6 & 7.0 & 0.337 & 0.054 & 0.243 & \underline{0.309} & 72.4 & 56.1 & 33.7 & 1.4\\
MatterGen  & 973 & \underline{99.6} & 74.5 & \textbf{85.7} & 6.9 & 0.209 & 0.019 & 0.302 & 0.451 & \underline{79.7} & \textbf{70.9} & \textbf{49.9} & \underline{3.9}\\
\midrule
SLayerGen3D & 91 & 99.5 & \underline{76.0} & 74.7 & \underline{9.9} & 0.870 & 0.055 & \underline{0.103} & \underline{0.309} & 66.9 & 53.3 & 27.5 & 2.1\\ \midrule
SLayerGen  & 78 & \textbf{99.9} & \textbf{76.9} & 70.7 & 7.5 & \underline{0.143} & \textbf{0.010} & 0.148 & \textbf{0.149} & 75.4 & \underline{63.9} & 34.2 & 2.1\\
SLayerGen ($T_{G,W}=2.0$)  & 79 & 99.5 & 71.0 & 80.6 & \textbf{20.0} & 0.253 & 0.021 & 0.192 & \textbf{0.212} & 65.8 & 52.6 & 32.3 & \textbf{4.4}\\
SLayerGen (Alex2D + MP20)  & 78 & \textbf{100.0} & \textbf{76.8} & 66.1 & 7.4 & \underline{0.145} & \textbf{0.011} & \textbf{0.093} & \textbf{0.131} & \textbf{80.6} & \underline{68.8} & 36.9 & 2.2\\\midrule
\end{tabular}}
\label{alex2d_results}
\vspace{-0.35cm}
\end{table}

%% file: evals_mp20.tex
\begin{table}[t!]
\centering
\caption{Results on the MP20 dataset of bulk triperiodic crystals.}
\vspace{-0.2cm}
\resizebox{\textwidth}{!}{
\begin{tabular}{ccccccccccccc}
\midrule
& \multicolumn{1}{c}{Time $\downarrow$} & \multicolumn{2}{c}{Validity (\%) $\uparrow$} & \multicolumn{2}{c}{U.N. rate (\%) $\uparrow$} & \multicolumn{3}{c}{Distribution distance $\downarrow$} & \multicolumn{2}{c}{Stable (\%) $\uparrow$}& \multicolumn{2}{c}{S.U.N. (\%) $\uparrow$}\\
 & (s / batch) & Structure & Composition & Structure & Template & $W_\rho$ & $W_{N_\mathrm{el}}$ & $\mathrm{JSD}_G$ & $\mathrm{E_{hull}}$ & $\mathrm{E_{hull}}\Gamma$ & Structure & Template\\ \midrule
SGEquiDiff   & \textbf{285} & 99.8 & 86.2 & 81.9 & 16.7 & \textbf{0.131} & 0.159 & 0.150 & 75.4 & 65.0 & 48.0 & 5.6\\
SLayerGen3D  & 318 & 99.8 & 85.2 & 82.4 & 17.9 & 0.204 & \textbf{0.124} & 0.151 & \textbf{76.7} & \textbf{65.1} & 46.6 & 6.0\\
SLayerGen (Alex2D + MP20)  & 331 & 99.8 & 84.4 & \textbf{89.5} & \textbf{24.9} & 0.297 & 0.188 & 0.153 & 69.0 & 59.4 & \textbf{48.4} & \textbf{7.8}\\ \midrule
\end{tabular}}
\label{mp20_results}
\vspace{-0.7cm}
\end{table}

%% file: section_conclusion.tex
\section{Conclusions}
In this paper, we proposed SLayerGen to generate crystals constrained by any space or layer group. We assembled several datasets and proposed evaluation metrics to facilitate progress in generative modeling of diperiodic materials; developed representations for layer groups, Wyckoff positions, and their associated asymmetric units; and built the first layer group invariant generative model, achieving state-of-the-art performance and enabling joint training on bulk and diperiodic materials.

\paragraph{Limitations and future directions} We used MLIP predictions as a proxy for DFT, which itself contains physical approximations. We did not explore how jointly training on tri/diperiodic materials might benefit from model scaling or architecture design. Our diperiodic materials stability metrics ignored effects from substrates \citep{ht_heterojunctions}, magnetism, and disorder. Building datasets and evaluations that consider these complexities, possibly with advancements in MLIPs \citep{magnetic_mace} and/or physical simulation \citep{dft_fe_precursor, dft_fe, quasicrystal_stability}, may help close the gap between \emph{in silico} design and experiment. Other interesting avenues for future research include post-training on symmetry-dependent properties, handling broken symmetries \citep{rpp_soft_equivariance}, and transfer learning across broad materials classes or properties \citep{uma, rees_moe}.

\paragraph{Impacts} This work has the potential to accelerate discovery of materials for energy, computing, optics, and more. Possible negative impacts include development of materials that require environmentally harmful processing or are used for military applications.

%% file: section_acknowledgements.tex
\begin{ack}
We are grateful for helpful discussions with Angela Pak, Barry Bradlyn, and Daniel Shoemaker. This research used the Illinois Campus Cluster, operated by the Illinois Campus Cluster Program in conjunction with the National Center for Supercomputing Applications. The research was supported by the National
Science Foundation under Grant Nos. DGE 21-46756 and 2118201.
\end{ack}

%% file: section_appendix.tex
\section{Appendix}

\input{orb_f1_scores}

\subsection{Examples of SUN templates}

    \begin{figure*}[h]
    \centering
    \includegraphics[width=\linewidth]{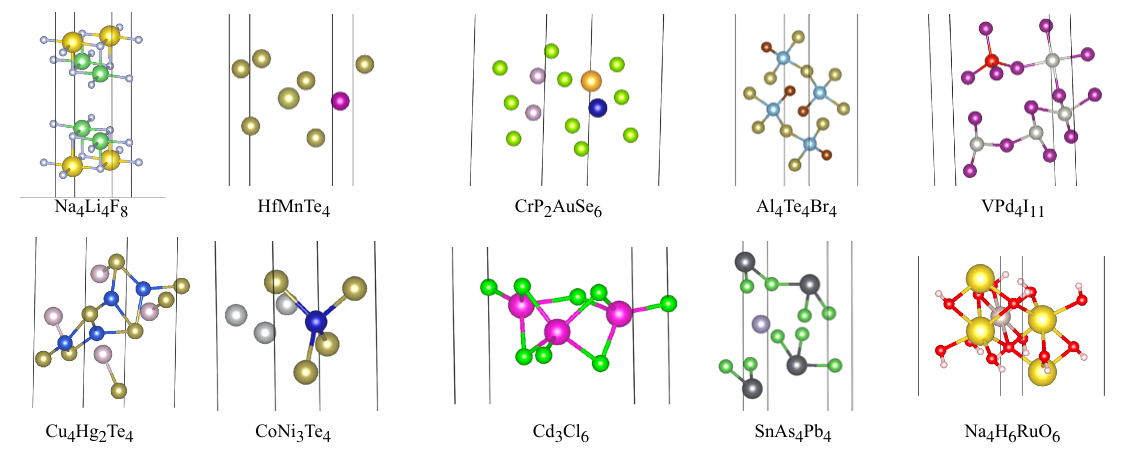}
    \vspace{-0.5cm}
    \caption{Example conventional unit cells of diperiodic materials generated by SLayerGen that passed our stability filters and had novel templates relative to the Alex2D training dataset. Some atoms have been translated outside the unit cell for visualization purposes.
    }
    \label{alex2d_sun_templates}
    \end{figure*}

\subsection{Space group lattice constraints}\label{space_group_lattices}
    For conventional unit cells of with lattice lengths $(a=||L_1||_2,b=||L_2||_2,c=||L_3||_2)$ and angles $(\alpha=\angle L_2L_3,\beta=\angle L_1L_3,\gamma=\angle L_1L_2)$, triclinic layer groups 1-2 impose no constraints, monoclinic oblique layer groups 3-7 require $\alpha=\beta=90^\circ$, monoclinic rectangular layer groups 8-18 require $\beta=\gamma=90^\circ$, orthorhombic layer groups 19-48 require $\alpha=\beta=\gamma=90^\circ$, tetragonal layer groups 49-64 require $a=b$ and $\alpha=\beta=\gamma=90^\circ$, and hexagonal layer groups 65-80 require $a=b$ and $\alpha=\beta=90^\circ$ and $\gamma=120^\circ$ \citep{layer_group_finding_algo}.
    
    Space groups 1 to 2 impose no constraints on the lattice; 3 to 15 require $\alpha=\gamma=90^\circ$; 16 to 74 require $\alpha=\beta=\gamma=90^\circ$; 75 to 142 require $\alpha=\beta=\gamma=90^\circ$ and $a=b$; 143 to 194 require $a=b$, $\alpha=\beta=90^\circ$, and $\gamma=120^\circ$; and 195 to 230 require $a=b=c$ and $\alpha=\beta=\gamma=90^\circ$. The entantiomorphic space groups require $|L|>0$ and consist of groups 76, 79, 91, 92, 95, 96, 144, 145, 151-154, 169-172, 178-181, 212, and 213.

\subsection{Removing redundancy from layer group Wyckoff positions}\label{layer_asus}
    We enumerated symmetrically unique regions of every Wyckoff position across the 80 layer groups. Since these can be found immediately for general Wyckoff positions as the interiors of the asymmetric units, we focus on the special Wyckoff positions which lie on asymmetric unit boundaries. For these, we first projected random points in $[0,1]^3$ onto special Wyckoff positions using projection matrices from \texttt{PyXtal} \citep{pyxtal} and then intersected the resulting points/lines/planes with the inexact ASUs from the International Tables of Crystallography \citep{itc_e} using \texttt{sympy}. To make these intersections \emph{exact}, we visually inspected the corresponding shapes in their asymmetric units using \texttt{PyVista} \citep{pyvista} and manually edited the shape vertices. For each Wyckoff position, we validated the completeness and uniqueness of the resulting shapes under layer group symmetry by projecting 1000 uniform random points in $[0,1]^3$ onto the Wyckoff position with \texttt{PyXtal} and asserting that each point's orbit lands within machine precision of our shape(s) exactly once.
    
\begin{figure*}[h]
    \centering
    \includegraphics[width=0.6\linewidth]{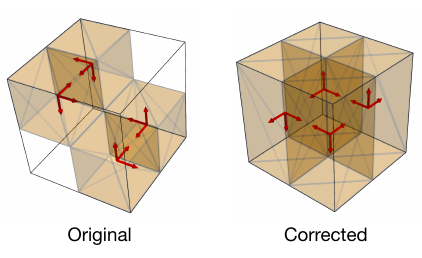}
    \vspace{-0.5cm}
    \caption{
    Left: The asymmetric unit (orange) listed by the International Tables of Crystallography for layer group 7, orbited by the layer group actions modulo lattice translations. We included an asymmetric object (red) at the center of each orbited asymmetric unit to visualize orientation. The conventional unit cell, outlined in black, is truncated to $z \in [-\frac{1}{2}, \frac{1}{2}]$ for visualization. The listed asymmetric unit does not tile $\mathbb{R}^3$. Right: Our asymmetric unit definition tiles $\mathbb{R}^3$.
    }
    \label{itc_corrections}
    \end{figure*}

    During processing of the Wyckoff positions, we discovered two errors in the International Tables of Crystallography \citep{itc_e}. First, we observed that the inexact asymmetric unit for layer group 7, listed as $0 \le x \le 1/2;\ 0 \le y \le 1/2; 0 \le z$, does not tile $\mathbb{R}^3$ upon application of the group actions. We corrected this asymmetric unit to $0 \le x \le 1/2;\ 0 \le y \le 1/2$, as shown in Fig. \ref{itc_corrections}. Second, we found that the inexact asymmetric unit for layer group 46 was listed as $0 \le x \le 1/4;\ 0 \le y \le 1/2$, which erroneously contained points from special Wyckoff position 4e in the asymmetric unit interior. Specifically, Wyckoff position 4e is defined at points $(x, \frac{1}{4}, z)\ \forall x,z \in \mathbb{R}$, which clearly resides in the asymmetric interior, for example, at $x=1/8$ and any $z$. By manual inspection, we corrected this inexact asymmetric unit to $1/4 \le x \le 3/4;\ 0 \le y \le 1/4$.

\subsubsection{Representing layer groups and their Wyckoff positions}\label{layer_group_wyckoff_embeddings}
    To encourage learning correlations between different space and layer groups, we modified the space group and Wyckoff position embeddings from SGEquiDiff \citep{sgequidiff} and extended them for layer groups. Specifically, space and layer groups were represented as the concatenation of 6-dimensional one-hot lattice centerings, 6-dimensional one-hot crystal families, 32-dimensional one-hot crystallographic point groups, a 1-dimensional indicator of whether the group supports chiral structures, a 1-dimensional indicator of whether the group is centrosymmetric, a 3-dimensional indicator of whether each lattice axis is periodic, and 16-dimensional arbitrary one-hot features to avoid collisions between groups resulting from having ignored screw and glide symmetries. This yielded $6+6+32+1+1+3+16=65$ total features per group. Each Wyckoff position was represented as the concatenation of the 65-dimensional crystallographic group embedding, 78-dimensional one-hot site symmetry group symbols, 4-dimensional indicators of how many degrees of freedom in the atomic coordinates are allowed by the Wyckoff position (i.e., 0-3), 17-dimensional one-hot Wyckoff multiplicities, and 48-dimensional mean-pooled Fourier embeddings of the centroids and vertices of the Wyckoff shapes, yielding $65+78+4+17+48=212$ features.

\subsection{Crystallographic Group Wrapped Normal}
    \paragraph{Proof sketch of score equivariance for layer groups}\label{layer_score_equivariance}

    For isotropic $\Sigma_t$ and orthogonal $R$, \citet{sgequidiff} proved that the score of the GWN is equivariant, i.e.,
    \begin{align}
        \nabla_{x_t}\log p(Rx_t + v|x_0;\Sigma_t) = R\nabla_{x_t}\log p(x_t |x_0;\Sigma_t).
    \end{align}
    This result also holds for the layer groups with our choice of anisotropic covariance $\Sigma_t = \sigma^2_{xy}I_2 \bigoplus \sigma^2_z I_1$. Specifically, since $R$ is block diagonal, i.e., $R=R^{xy} \bigoplus R^z$ for $R^{xy} \in O(2)$ and $R^z \in O(1)$, then the proof of \citet{sgequidiff} directly applies to each subspace independently.
    
    \paragraph{Non-orthogonality in fractional coordinates for hexagonal groups}\label{hexagonal_scores}
    While all crystallographic groups strictly contain isometries in real space, when working in fractional coordinates, hexagonal groups contain group elements $g=\{R_g, v_g\}$ where $R_g \notin O(3)$. These groups have conventional lattice parameter constraints following $a=b$, $\alpha=\beta=90^\circ$, and $\gamma=120^\circ$, yielding lattice matrices of the form
    \begin{align*}
        L=
        \begin{pmatrix} 
            a & -a/2 & 0\\
            0 & a\sqrt{3}/2 & 0\\
            0 & 0 & c
        \end{pmatrix}
    \end{align*}
    where each column is a lattice vector. Naively, using $L$, we could compute our Group Wrapped Normal in the Cartesian basis where group representations are orthogonal. However, we wish for scores of the Group Wrapped Normal to be invariant to the lattice size so that we can choose a single noise scheduler for all crystals. Thus, for hexagonal groups, we choose the following canonical basis:
    \begin{align*}
        B=
        \begin{pmatrix} 
            1 & -1/2 & 0\\
            0 & \sqrt{3}/2 & 0\\
            0 & 0 & 1
        \end{pmatrix}.
    \end{align*}
    In this basis, given $g=\{R,v\}$ where $R$ and $v$ act on fractional positions $x\in\mathbb{R}^3$, group actions become
    \begin{align*}
        B(Rx + v) &= (BRB^{-1})(Bx) + Bv\\
            &\equiv \tilde{R}\tilde{x} + \tilde{v}
    \end{align*}
    where $\tilde{R}=BRB^{-1}\in O(3)$, $\tilde{x}=Bx$, and $\tilde{v}=Bv$. We compute scores $\nabla_{\tilde{x}_t}\log p(\tilde{x}_t|\tilde{x}_0)$ the same way as SGEquiDiff \citep{sgequidiff} but plug in our transformed variables:
    \begin{align*}
        \nabla_{\tilde{x}_t}\log p(\tilde{x}_t|\tilde{x}_0) &= \frac{\sum\limits_{\{R_j,v_j\}\in G} \exp\Big(\frac{-[\tilde{x}_t - (\tilde{R}_j \tilde{x}_0 + \tilde{v}_j)]^T \Sigma_t^{-1} [\tilde{x}_t - (\tilde{R}_j \tilde{x}_0 + \tilde{v}_j)]}{2}  \Big)\Sigma_t^{-1}(\tilde{R}_j \tilde{x}_0 + \tilde{v}_j - \tilde{x}_t)}{p(\tilde{x}_t|\tilde{x}_0)}\\
        &= \frac{\sum\limits_{\{R_j,v_j\}\in G} \exp\Big(\frac{-[B(x_t - (R_j x_0 + v_j))]^T \Sigma_t^{-1} [B(x_t - (R_j x_0 + v_j))]}{2}  \Big)\Sigma_t^{-1} B(R_j x_0 + v_j - x_t)}{\sum\limits_{\{R_i,v_i\}\in G} \exp\Big(\frac{-[B(x_t - (R_i x_0 + v_i))]^T \Sigma_t^{-1} [B(x_t - (R_i x_0 + v_i))]}{2}\Big)}.
    \end{align*}
    Finally, we transform the score back to fractional space as $B^{-1}\nabla_{\tilde{x}_t}\log p(\tilde{x}_t|\tilde{x}_0)$.
    
    Using the equivariance property, i.e., $\nabla_{\tilde{x}_t} \log p(\tilde{R}\tilde{x}_t + \tilde{v}|\tilde{x}_0) = \tilde{R} \nabla_{\tilde{x}_t}\log p(\tilde{x}_t | \tilde{x}_0)$, and defining the induced action of $g=\{R,v\}$ on the fractional score as
    \begin{align*}
        \phi(g\cdot x_t|x_0) &\equiv B^{-1}\nabla_{Bx_t}\log p(B(g\cdot x_t)|Bx_0)\\
        &= B^{-1}\nabla_{Bx_t}\log p(B(Rx_t + v)|Bx_0)\\
        &= B^{-1}\nabla_{Bx_t}\log p(\tilde{R}x_t + \tilde{v}|Bx_0),
    \end{align*}
    we see 
    \begin{align*}
    B^{-1}\nabla_{\tilde{x}_t}\log p(\tilde{R}\tilde{x}_t + \tilde{v}|\tilde{x}_0) &= B^{-1}\tilde{R}\nabla_{\tilde{x}_t}\log p(\tilde{x}_t |\tilde{x}_0) \\
         &= RB^{-1} \nabla_{\tilde{x}_t}\log p(\tilde{x}_t |\tilde{x}_0)\\
        \implies \phi(g \cdot x_t|x_0) &= R\phi(x_t|x_0),
    \end{align*}
    showing $\phi(\cdot)$ has the same desired equivariance behavior as GWN scores for non-hexagonal groups and symmetrized functions $f(\cdot)$ (Eq. \ref{sg_equivariance}) that \citet{space_group_equivariant_neural_networks} showed satisfies $f(g\cdot x)=R f(x)$.

\subsection{Diffusion training and inference}\label{diffusion_details}
    We followed prior crystal diffusion models \citep{diffcsp, diffcsp_pp, sgequidiff, mattergen} and used a variance-exploding log-uniform noise scheduler $\log\sigma_t \sim \mathcal{U}(\log \sigma_0, \log \sigma_T)$, predictor-corrector stochastic sampling \citep{sde_diffusion} with a signal-to-noise ratio of 0.4, $T=1000$ diffusion steps, and $\sigma_{0,xy} = \sigma_{0,z} = 0.002$. For periodic coordinates, we set $\sigma_{T,x_\mathrm{periodic}}=0.5$. For aperiodic coordinates, we set $\sigma_{T,x_\mathrm{aperiodic}}=67.0$ for BiDB and $\sigma_{T,x_\mathrm{aperiodic}}=45.0$ for 2DMatPedia, C2DB, and Alex2D.

    We trained with the score matching loss,
    \begin{align*}
        L_X &= \mathbb{E}_{x_t \sim p(x_t | x_0, W), x_0 \sim p_\mathrm{data}, t \sim \mathcal{U}(0,T)}[\\
        &\omega_\mathrm{periodic}||s_\theta(x_t, t) \odot \mathds{1}_\mathrm{periodic} - \lambda_t \nabla_{x_t} \log p(x_t | x_0) \odot\mathds{1}_\mathrm{periodic}||^2_2
        \\&+ \omega_\mathrm{aperiodic}||s_\theta(x_t, t) \odot \mathds{1}_\mathrm{aperiodic} - \sigma_{t,z} \nabla_{x_t} \log p(x_t | x_0)  \odot\mathds{1}_\mathrm{aperiodic}||^2_2\\
        ],
    \end{align*}
    where $\odot$ is elementwise multiplication, $\mathds{1}_\mathrm{aperiodic} = \mathbf{1}-\mathds{1}_\mathrm{periodic} \in \mathbb{Z}_2^3$ is 1 for aperiodic dimensions and 0 for periodic ones, and loss weight $\lambda_t = \mathbb{E}_{x_t \sim p(x_t|x_0), x_0\sim p_{w,\mathrm{prior}}}[||\nabla_{x_t}\log p(x_t|x_0) \odot \mathds{1}_\mathrm{periodic}||]^{-1} \in \mathbb{R}$. We set $\omega_\mathrm{periodic}=2$ and $\omega_\mathrm{aperiodic}=1$ in our experiments.
    
    Noisy samples $x_t$ were created by sampling Gaussian noise $\epsilon \sim \mathcal{N}(0, I)$ and reparameterizing as $x_t = x_0 + \mathcal{P}_w(\sigma_t \epsilon)$ where $\mathcal{P}_w(\cdot)$ is an orthogonal projection to the tangent space of the Wyckoff position $w$ of $x_0$. We set $p_{w, \mathrm{prior}}$ to uniform $\mathcal{U}[0,1] \cap w$ for periodic coordinates and Gaussian $\mathcal{N}(0, \sigma^2_{t,z})$ for aperiodic coordinates.
    
    To construct the uniform distribution over Wyckoff sites $\mathcal{U}[0,1] \cap w$, we sampled uniformly on the Wyckoff shapes. Specifically, we sampled a point for 0D Wyckoff positions. For 1D Wyckoff positions, we sampled a line segment proportionally to its length and then uniformly in the segment. For 2D Wyckoff positions, we pre-computed Delaunay triangulations of the polygonal facets belonging to the Wyckoff position, sampled a triangle proportionally to its surface area, and then sampled in the triangle with a uniform Dirichlet distribution. Coordinates in 3D Wyckoff positions were sampled with rejection sampling from a bounding box around the ASU.
    
    Loss weights for periodic coordinates were approximated with Monte Carlo sampling with $s$ samples as
    \begin{align*}
        \lambda_t \approx \Big[\frac{1}{s} \sum_{i=1}^s \nabla_{x_t^i} \log p(x_t^i | x_0^i; G, w, \Sigma_t)  \Big]^{-1}
    \end{align*}
    In our experiments, we set $s=2500$. Since crystallographic groups consist of an infinite number of lattice translations, we cannot compute the score of the wrapped normal explicitly. Instead, following prior work \citep{diffcsp, diffcsp_pp}, we approximated it with a truncated sum over lattice translations as
    \begin{align*}
        &q(x_t|x_0) \propto \\
        &\sum_{t_L \in \mathbb{Z}^d \cap [-m, m]^d} \sum_{\{R_j, v_j\} \in G/T_L} \exp\Big[\frac{-1}{2}(x_t - R_jx_0 - v_j - t_L)^T\Sigma_t^{-1}(x_t - R_jx_0 - v_j - t_L)) \Big]
    \end{align*}
    where $d=2$ for layer groups and $3$ for space groups. We used $m=3$.

\subsection{Model architecture}
    \subsubsection{Wyckoff-Element Transformer}
        We employed the transformer architecture in SGEquiDiff but leveraged our layer group and Wyckoff position embeddings (Sec. \ref{layer_group_wyckoff_embeddings}):
        \begin{align*}
            z^0 &\leftarrow \mathrm{MLP}(e_{G} || e_W || e_A)\\
            z^{l+1} &\leftarrow \mathrm{EncoderLayer}(z_i^l, m_\mathrm{causal})\\
            z_W, z_A &\leftarrow \mathrm{Split}(z^{l_\mathrm{max}})\\
            z_A &\leftarrow \mathrm{MLP}(z_A, e_W)\\
            p_{W,\mathrm{stop}} &\leftarrow \mathrm{MultiheadDotProductAttentionScore}(K=[e^\mathrm{all}_W||e_\mathrm{stop}], Q=z_W, \mathrm{mask}=m_W)\\
            p_A &\leftarrow \mathrm{MultiheadDotProductAttentionScore}(K=e^\mathrm{all}_A, Q=z_A, \mathrm{mask}=m_A)
        \end{align*}
        where $e_G$, $e_W$, and $e_A$ are embeddings of the crystal's space or layer group, occupied Wyckoff positions, and atomic elements, respectively; $l_\mathrm{max}$ is the number of encoder layers; $m_\mathrm{causal}$ is a causal attention mask enforcing atom orderings; $e^\mathrm{all}_W$, $e_\mathrm{stop}$, and $e^\mathrm{all}_A$ are predicted embeddings of all sampleable Wyckoff positions, the stop token, and all sampleable elements, respectively; $m_W$ and $m_A$ are attention masks enforcing lexicographic atom orderings; $p_{W,\mathrm{stop}}$ and $p_A$ are model probabilities over Wyckoff positions, the stop token, and elements; and $\mathrm{EncoderLayer(x, m)}$ is summarized by the following:
        \begin{align*}
            x_\mathrm{norm} &\leftarrow \mathrm{LN}(x)\\
            x &\leftarrow x + \mathrm{MultiheadSelfAttention}(x_\mathrm{norm}, m)\\
            x &\leftarrow \mathrm{Dropout}(x)\\
            x &\leftarrow \mathrm{Dropout}\big(x + \mathrm{Linear(Dropout(MLP(}x))) \big)
        \end{align*}
    
    \subsubsection{Graph Neural Network}\label{gnn_architecture}
        We modified the FAENet-inspired GNN \citep{faenet} from SGEquiDiff \citep{sgequidiff} to be configurably invariant to either P1 space or layer groups such that it can be symmetrized to any space or layer group with Eq. \ref{sg_equivariance}. We used fully-connected graphs in the conventional unit cell with edge features based on relative positions in fractional coordinates (where aperiodic fractional coordinates are scaled to unit variance for stability) and distances in real space. For the real space distances, we used the minimum image convention, where images were masked out from the aperiodic dimension for layer groups:
        \begin{align*}
            r_{ij} = \min_j r(x_i, x_j + n_1e_1 + ... + n_d e_d).
        \end{align*}
        Here, $x_i$ is in fractional coordinates, $n_i \in \mathbb{Z}$, $e_i \in \mathbb{Z}^3$ is an integer lattice translation, $d\in[2,3]$ indicates how many periodic dimensions there are, and $r(\cdot, \cdot)$ is the distance function.
        
        To embed (relative) positions in fractional coordinates, we used Fourier features. For periodic coordinates, we sampled $n_\mathrm{freq}$ integer frequencies $\nu_{k,\mathrm{periodic}} \in \mathbb{Z}^3 \cap [-512,512]^3 \setminus \mathbf{0}$ without replacement with probabilities drawn from a discretized standard normal distribution. For aperiodic coordinates, we sampled $n_\mathrm{freq}$ random frequencies $\nu_{k,\mathrm{aperiodic}} \sim \mathcal{N}(0,1) \in \mathbb{R}$. The architecture is summarized as follows:
        \begin{align*}
            x_i &\leftarrow (x_i\bmod 1) \odot \mathds{1}_\mathrm{periodic} + (x_i -\frac{1}{N}\sum_{i=1}^Nx_i)\odot \mathds{1}_\mathrm{aperiodic} \\
            t_\mathrm{emb} &= \mathrm{MLP}(c_\mathrm{noise}(\mathrm{diag}(\Sigma_t)))\\
            e_d &= \mathrm{MLP}(\mathds{1}_\text{Crystal has space group symmetry})\\
            h_i^0 &\leftarrow \mathrm{MLP}\big(e_{a_i}|| \mathrm{FourierEmbedding}(x_i, \Sigma_{t,33})\big || t_\mathrm{emb} || e_d)\\
            e_{ij} &\leftarrow \mathrm{MLP\big(\mathrm{FourierEmbedding}(x_j - x_i, \Sigma_{t,33})|| \mathrm{RBF}(r_{ij}) || \mathrm{Norm}(\mathbf{l}) || e_d \big)}\\
            f_{ij}^l &\leftarrow \mathrm{MLP}\big(e_{ij}||h_i^l||h_j^l\big)\\
            h_i^{l+1}&\leftarrow h^l_i+a\cdot \mathrm{MLP\circ\mathrm{GraphNorm}\Bigg(\frac{1}{\sqrt{|\mathcal{N}_i|}}\sum\limits_{j\in\mathcal{N}_i} h_j^l \odot f_{ij}^l\Bigg)}\\
            h_i^{l_\mathrm{max}} &\leftarrow \mathrm{MLP}(h_i^0||...||h_i^{l_\mathrm{max}})\\
            \hat{f}_i &\leftarrow \mathrm{MLP}(h_i^{l_\mathrm{max}})
        \end{align*}
        where
        \begin{align*}
            c_\mathrm{noise}(\cdot) &= \frac{1}{4}\log(\cdot)\\
            c_\mathrm{in}(\cdot) &= 1 / \sqrt{(\cdot)^2 + \sigma_{z,\mathrm{data}}^2}\\
            \mathrm{FourierEmbedding(x \in \mathbb{R}^3, \sigma_{t,z})} &\leftarrow \big[\\
            \mathds{1}_\mathrm{crystal\ is\ diperiodic} &||c_\mathrm{in}(\sigma_{t,z}) x_z \mathds{1}_\mathrm{crystal\ is\ diperiodic} ||\big(\cos(2 \pi x^T w_i) || \sin (2\pi x^T w_i)\big)_{i=0,...,n_\mathrm{freq}}\\
            \big]\\
            w_i &= 2\pi \cdot \begin{cases}
                \nu_{i,\mathrm{periodic}} & \text{if } d=3 \\ 
                (\nu_{i,\mathrm{periodic}}[:2]|| \nu_{i,\mathrm{aperiodic}}) & \text{otherwise} 
            \end{cases} \in \mathbb{R}^3.
        \end{align*}
        
        Here, $N$ is the number of atoms in the conventional unit cell, $\mathds{1}_\mathrm{periodic}\in \mathbb{Z}_2^3$ is an indicator vector that is 1 for periodic coordinates and 0 for aperiodic ones, $\mathcal{N}_i$ are the neighbors of atom $i$, $a$ is a learnable scalar initialized at zero, $\mathrm{RBF}(d_{ij})$ featurizes Cartesian distances in log space using a basis of 50 Gaussians with means from a log-uniform grid between $\log(10^{-6})$ and $\log(\sigma_{T,z})$, $\mathbf{l}=(a,b,c,\alpha,\beta,\gamma)$ are the conventional unit cell lattice parameters where we masked $c=-1$ for layer group symmetric crystals, $\mathrm{Norm}(\cdot)$ applies min-max normalization,  $\mathrm{GraphNorm}$ \citep{graphnorm} applies featurewise Z-score normalization over nodes with learnable shift and scaling parameters, $\sigma_{z,\mathrm{data}}^2$ is the dataset variance over aperiodic fractional coordinates, and $c_\mathrm{noise}$ and $c_\mathrm{in}$ are employed for training stability \citep{karras_edm}.

        We see that $\hat{f}_i$ is invariant to lattice translations because its spatial inputs are minimum-image Cartesian distances under periodic boundary conditions and fractional coordinates with periodic dimensions featurized by plane waves. 
        
\subsection{Model hyperparameters}
    Our model was trained with the Adam optimizer \citep{adam} on a single NVIDIA A40 GPU. Single-task training on the largest diperiodic materials dataset (Alex2D) took approximately 9 hours. Model evaluation on each dataset took 1-2 hours with a single A40 GPU and 20 CPU cores. Hyperparameters were tuned by manually. The lattice sampler, Wyckoff-element transformer, and diffusion model were trained simultaneously but independently. Early stopping based on validation loss was applied on a per-module basis.

    \begin{table}[h]
    \begin{center}
    \begin{tabular}{ll}
    \multicolumn{1}{c}{\bf Hyperparameter}  &\multicolumn{1}{c}{\bf Value}
    \\ \hline \\
    Batch size & 256\\
    Number of epochs & 2000\\
    Lattice length noise range & 0.3\\
    Lattice angle noise range & 5.0\\
    Lattice sampler hidden dimension & 256\\
    Lattice length Fourier scale & 5.0\\
    Lattice angle Fourier scale & 1.0\\
    Lattice length bin edges Fourier scale & 2.0\\
    Lattice angle bin edges Fourier scale & 1.0\\
    Lattice parameter embedding dimension & 32\\
    Lattice sampler number of hidden layers & 3\\
    Transformer dropout rate & 0.1\\
    Transformer hidden dimension & 256\\
    Transformer number of hidden layers & 4\\
    Transformer number of heads & 2\\
    GWN lattice translations & 3\\
    GWN Monte Carlo samples & 2500\\
    Time embedding dimension & 128\\
    Number of plane wave frequencies & 96\\
    Number of Cartesian distance radial basis functions & 96\\
    Edge embedder hidden dimension & 128\\
    Node embedder hidden dimension & 256\\
    Number of message passing steps & 5\\
    Lattice learning rate & $1\times 10^{-4}$\\
    Transformer learning rate & $1\times10^{-4}$\\
    GNN learning rate & $1\times10^{-3}$\\
    Weight decay & 0.0\\
    Learning rate scheduler & None\\
    Gradient clipping by value & 0.5\\
    Joint Alex2D/MP20 training - Alex2D sampling weight & 1.3\\
    Joint Alex2D/MP20 training - MP20 sampling weight & 1.0\\
    \midrule
    \end{tabular}
    \end{center}
    \end{table}

\subsection{Dataset details}\label{dataset_settings}
2DMatPedia and Alexandria are released under the \href{https://creativecommons.org/licenses/by/4.0/}{CC BY 4.0 license}. C2DB and BiDB are released under the \href{https://creativecommons.org/licenses/by-nc/4.0/}{CC BY-NC 4.0 license}. For full details on each dataset's DFT settings, please refer to the corresponding papers. Here, we briefly summarize them:
\begin{itemize}
    \item 2DMatPedia \citep{2dmatpedia}: VASP vdW-optB88 functional with the Materials Project's +U corrections \citep{materialsproject}, 520 eV plane wave cutoff, spin polarization, 1e-4 eV ionic convergence, Materials Project $k$-point densities, 20\AA\ vacuum
    \item C2DB \citep{c2db}: GPAW PBE functional, 800 eV plane wave cutoff, 0.01 eV/Angstrom force convergence, 6\AA$^{-1}$ $k$-point densities, 15\AA\  vacuum, 0.05 eV Fermi smearing
    \item BiDB \citep{bidb}: GPAW PBE-D3 with custom +U corrections, 800 eV plane wave cutoffs, spin polarization, 12\AA$^{-1}$\ $k$-point densities, 15\AA\  vacuum, 50 meV Fermi-Dirac smearing
    \item Alexandria \citep{alexandria}: VASP v5.2 PBE+U, 520 eV cutoff, $5\times 10^{-3}$ eV/A force convergence, 15\AA\ vacuum
\end{itemize}

\subsubsection{Layer centering and monolayer isolation from bilayers}\label{layer_centering}
    Since raw structures from 2DMatPedia, C2DB, BiDB, and Alexandria were represented with a bounded 3D periodic unit cell, the superficial periodic boundaries along the aperiodic $c$-axis often ``cut'' layers. To make atoms always appear contiguously, we checked consecutive differences in each atom position's fractional $z$ coordinate. If the largest difference was >7\AA, we determined that vacuum was found in the middle of the cell (and thus the layer was erroneously cut by the periodic boundary) and translated all atoms along the $c$-axis until they wrapped contiguously to the bottom of the cell. The atoms were then translated to $z=0.5$. 

    Similarly, to isolate monolayers from bilayers for computing binding energies and interlayer distances, we found the largest gap in consecutive $z$-coordinates. All atoms above and below the gap were classified as belonging to the upper and lower monolayer, respectively.

\begin{figure*}[p]
\centering
\includegraphics[width=\linewidth]{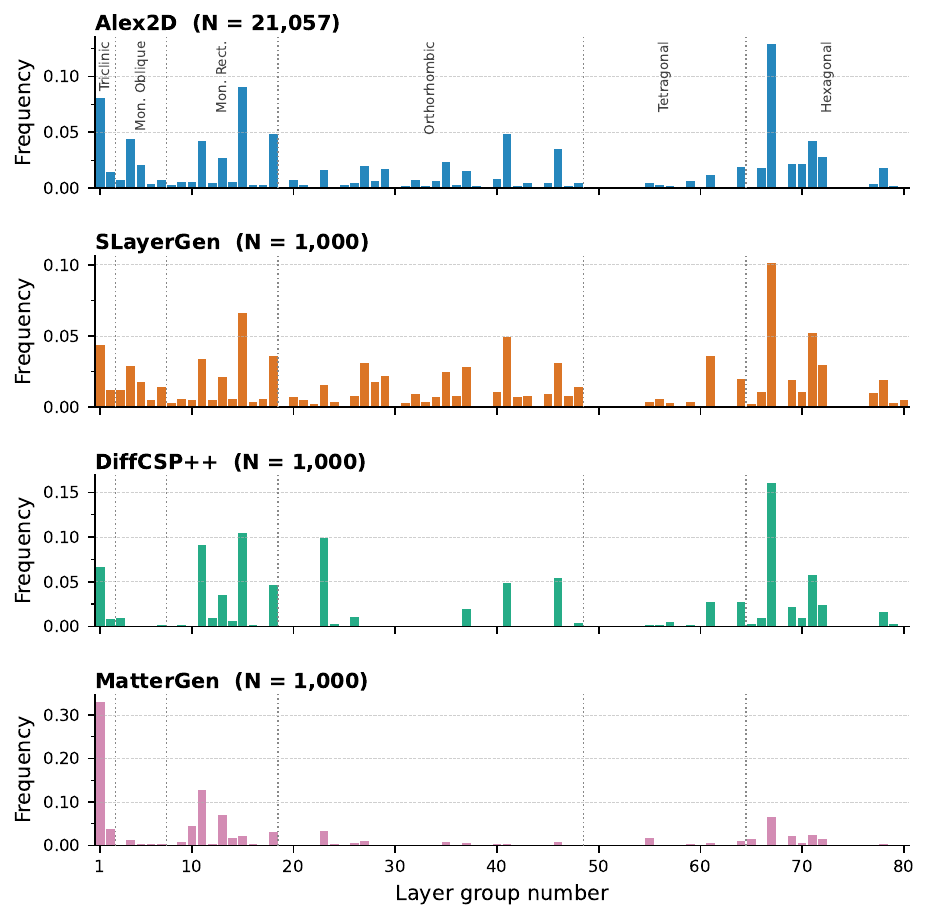}
\vspace{-0.5cm}
\caption{
Histograms of occupied layer groups by samples from generative models and by their training data from Alexandria \citep{alexandria, alexandria_2d_strategy}. Layer groups were determined by the \texttt{get\_layergroup} function in \texttt{spglib} \citep{layer_group_finding_algo, spglib} using tolerances of 0.01\AA\ for ground-truth DFT-relaxed structures and 0.1\AA\ for samples from generative models. Vertical dashed lines separate layer groups by crystal system. We observe that SLayerGen preserves the layer group distribution of the training set, DiffCSP++ undersamples many orthorhombic and monoclinic groups, and MatterGen heavily oversamples low symmetry P1 crystals.}
\label{all_model_layer_hists}
\end{figure*}

\clearpage
\begin{figure*}[p]
\centering
\includegraphics[width=\linewidth]{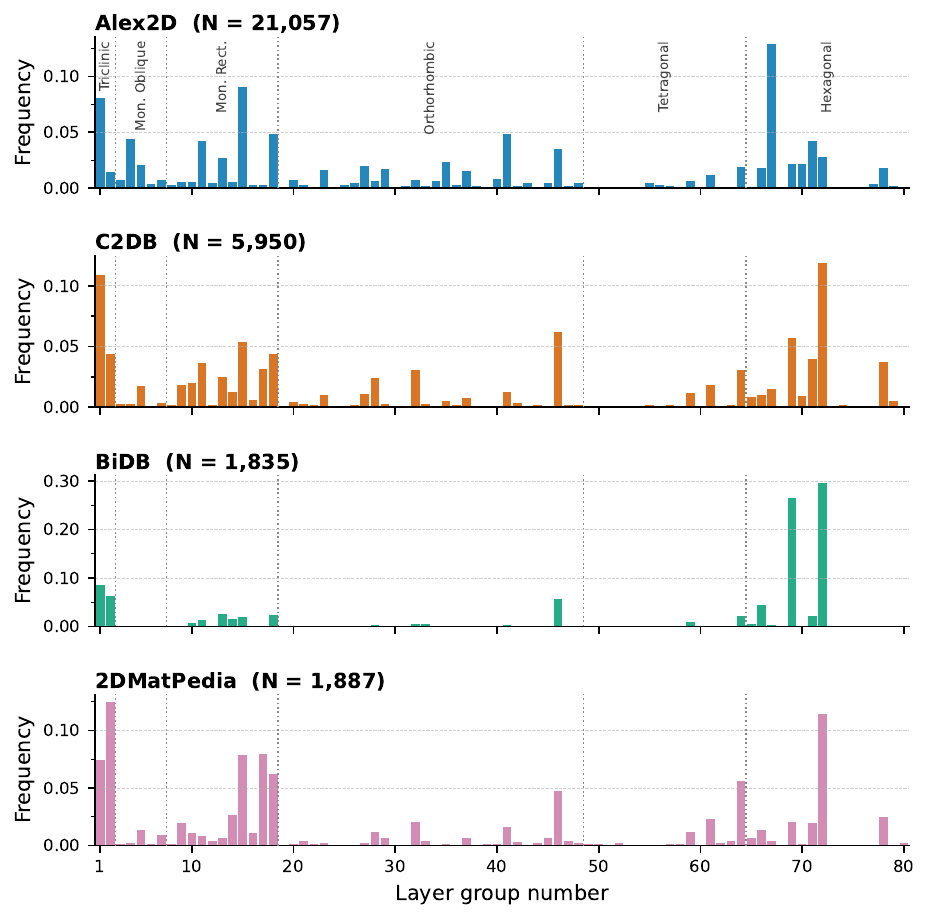}
\vspace{-1cm}
\caption{
Histograms of occupied layer groups by our filtered training data from Alexandria \citep{alexandria, alexandria_2d_strategy}, C2DB \citep{c2db}, BiDB \citep{bidb}, and 2DMatPedia \citep{2dmatpedia}. Layer groups were determined by the \texttt{get\_layergroup} function in \texttt{spglib} \citep{layer_group_finding_algo, spglib} using tolerances of 0.01\AA. Vertical dashed lines separate layer groups by crystal system.}
\label{all_dataset_layer_hists}
\end{figure*}
\clearpage

%% file: orb_f1_scores.tex

\begin{table}[h]
\centering
\caption{F1 scores for Orb-v3 using Materials Project \texttt{v2023.02.07} convex hulls on 2,000 random data samples and for a dummy classifier predicting everything as stable on full unfiltered datasets. Stability was defined as having an energy above the hull less than 200 meV/atom. We excluded evaluation on the BiDB dataset because it only contains bilayers with stable binding energies, rendering F1 scores meaningless.}
\label{orb_f1_scores}
\vspace{0.2cm} 
\small 
\begin{tabular}{lcccccc} 
\toprule
& \multicolumn{2}{c}{2DMatPedia} & \multicolumn{2}{c}{C2DB} & \multicolumn{2}{c}{Alex2D} \\
\cmidrule(lr){2-3} \cmidrule(lr){4-5} \cmidrule(lr){6-7}
& Orb-v3 & Dummy & Orb-v3 & Dummy & Orb-v3 & Dummy \\ 
\midrule
F1 Score & 0.914 & 0.723 & 0.903 & 0.726 & 0.825 & 0.480 \\
\bottomrule
\end{tabular}
\end{table}